\documentclass[smallextended]{svjour3} 
\smartqed
\usepackage{graphicx}
\usepackage{amsmath}
\usepackage{verbatim}
\begin{document}

\title{A fractional order recovery SIR model from a stochastic process}

\author{C. N. Angstmann \and B. I. Henry \and \\A. V. McGann}
\institute{C. N. Angstmann \at
              School of Mathematics and Statistics, UNSW Australia, Sydney 
Australia 2052 \\
              \email{c.angstmann@unsw.edu.au}    
           \and
          B. I. Henry \at
              School of Mathematics and Statistics, UNSW Australia, Sydney 
Australia 2052 \\
              \email{b.henry@unsw.edu.au}    
                         \and
          A. V. McGann \at
              School of Mathematics and Statistics, UNSW Australia, Sydney 
Australia 2052 \\
              \email{a.mcgann@unsw.edu.au}    
}

\date{Received: date / Accepted: date}
\maketitle

\begin{abstract}

Over the past several decades there has been a proliferation of epidemiological models with ordinary derivatives replaced by fractional derivatives in an {\em an-hoc} manner. These models may be mathematically interesting but their 
relevance is uncertain. 
Here we develop an SIR model for an epidemic, including vital dynamics,  from an  underlying stochastic process.
 We show  how fractional 
differential operators arise naturally in these models whenever the recovery time from the disease is power law distributed.  
This can provide a model for a chronic disease process where individuals who are infected for a long time are unlikely to recover.
The fractional order recovery model is shown to be consistent with the 
 Kermack-McKendrick age-structured SIR model and it reduces to the Hethcote-Tudor integral equation SIR model. 
The derivation from a stochastic process is extended to discrete time, providing a stable numerical method for solving the model equations.

We have carried out  simulations of the fractional order recovery model
showing convergence to equilibrium states.
The number of infecteds in the endemic equilibrium state increases as the fractional order of the derivative tends to zero.


\end{abstract}

\section{Introduction}

The classic SIR model was introduced by Kermack and McKendrick in 1927 \cite{KM1927}. In the simplest form of this model the population is separated into three compartments representing susceptible (S), infected (I) and removed (R) populations. The time evolution of the disease is then described by three coupled ordinary differential equations. Kermack and McKendrick also considered more general age-structured SIR models in 1932 and 1933 \cite{KM1932,KM1933}. These more general models were formulated as a set of coupled integro-differential equations. Numerous other variants of the SIR model, and related models, have been widely studied over the past several decades \cite{H2000}. The classical SIR models have also been formulated as the expectation of stochastic processes \cite{AB2000}. 

 It is easy to find numerous studies in the literature of {\em ad-hoc} fractional order SIR models where the time derivatives are simply replaced with  nonlocal Caputo fractional derivatives
 \cite{AAD2015,AR2012,AE2013,DU2011,D2013,DH2009,GG2014,GM2014,OD2011,PR2011,ZZ2013}. These models, and their analysis, are mathematically interesting, however they have no particular physical motivation, other than the incorporation of a memory effect. 
 Spatially dependent SIR models with space fractional derivatives have also been considered \cite{HSD2011} to account for nonlocal spread of disease.

In this work we show that  fractional order time derivatives can be included in the governing equations for an SIR model derived from
a physical underlying stochastic process.
We consider a variation of an SIR model where recovery from the disease is dependent on the time since infection. 
The model is derived from a directed continuous time random walk through the SIR compartments, with the time in the infectious 
compartment drawn from a waiting time probability density. 
We show that, in the case of a power law tailed waiting time density, the governing equations become a set of fractional order differential equations. 
The expected recovery time diverges in a power law waiting time density and this
 leads to chronic infection in the fractional order SIR model. As the fractional order derivative operates on the 
 recovery we refer to this as the frSIR model.
 Other fractional epidemic compartment models, such as a fractional SEIR model,  could also be derived following the approach outlined here. There have been several studies of semi-Markovian epidemic models in the recent literature \cite{KL2007,MH2011} that are related to the approach presented here, but they are not formulated as coupled integro-differential equations.

Starting with a discrete time stochastic process formulation of the frSIR model we
 derive a numerical method for solving the governing fractional order differential equations.
 The numerical method is related to the discrete time stochastic process method that was recently introduced to
solve the fractional Fokker-Planck equation \cite{ADHN2014}. We have implemented the numerical scheme to investigate the effects of changes in the fractional order exponent on the qualitative behaviour of solutions. 
The numerical solutions converge to the calculated equilibrium states of the frSIR model, when the parameters are constants.

In section \ref{sec_derivation} we derive an SIR model with an arbitrary waiting time before transitioning from the infectious compartment to the recovered compartment. This general model is shown to be consistent  with the
structured formulation of Kermack and McKendrick \cite{KM1932,KM1933}. We also derive an integral equation representation of the model and show that it reduces to the 
 integral equations presented by Hethcote and Tudor \cite{HT1980} when the parameters are constants.
In section \ref{sec_fSIR} we show that in the case of a power-law waiting time distribution in the infectious compartment we obtain fractional order derivatives in the model and we present the governing equations for the frSIR model. 
In section \ref{sec_numerical} a stable numerical scheme for solving the frSIR model is derived from a discrete time formulation of the stochastic process. In Appendix B
 the discrete time formulation is shown to converge, under a continuous time limit, to the continuous time formulation. Numerical solutions are investigated in
section \ref{sec_example}, and we conclude with a summary in section \ref{sec_conclusion}.

\section{SIR as a Continuous Time Random Walk}
\label{sec_derivation}

An underlying assumption in the simplest SIR models is that the transition of an individual through each of the compartments is independent of the amount of time since the individual entered the compartment. This assumption is mathematically equivalent to assuming that the time spent in each compartment is exponentially distributed. This is a very restrictive assumption with no a priori reason for it to hold. In some diseases with the potential for chronic infection, such as human papillomavirus (HPV), there is evidence of power-law tails in the distribution of infected times \cite{Rositch2013}. We have incorporated an arbitrary time in the infected compartment in our derivation of a generalised SIR model below by way of a continuous time random walk (CTRW) \cite{MW1965,SL1973}.

In general our derivation may be adapted to any compartment model where the transition out of a compartment is dependent on the length of time since entering the compartment. In this work we have concentrated on an SIR model with births and deaths.  

In the standard manner, we separate the population into three compartments, Susceptible (S),  Infectious (I), and Recovered (R) \cite{H2000}. The population is composed of individuals who are born into the S compartment and undergo a directed CTRW on the S, I, and R, compartments until they die and are removed from consideration. As in the standard model, individuals may only move from the S compartment to the I compartment and then to the R compartment. The transition to the I compartment occurs when an individual becomes infected and the transition to the R compartment occurs when an individual recovers from the infection. The derivation of fractional diffusion equations \cite{HA1995,MK2000},
fractional reaction diffusion equations \cite{HLW2006,SSS2006,F2010},
fractional Fokker-Planck equations \cite{BMK2000,SK2006,HLS2010} and fractional
chemotaxis diffusion equations \cite{LH2010,F2011} from CTRWs has been 
well studied and provides clear physical motivation for each of these systems. 
Our derivation of the evolution equations for the SIR model and fractional SIR model from a stochastic CTRW process with reactions is similar to the derivation of the fractional Fokker-Planck equation with reactions \cite{ADH2013mmnp} and the derivation of the master equations for CTRWs with reactions on networks \cite{ADH2013pre}.

Consider an individual who is infectious. The probability that they will 
infect a susceptible person in the time $t$ to $t+\delta t$ is assumed to be a 
product of the probability that the infectious person will encounter a 
susceptible and the probability that an encounter with a susceptible will result in an infection. Without loss of generality we can express  the probability that
an encounter with a susceptible will result in infection by $\omega(t) \delta t+o(\delta t)$, identifying $\omega(t)$ as the rate of becoming infected per time interval $\delta t$. Given that there are $S(t)$  susceptible people at time $t$, this implies that the probability of an infected individual creating a new infected individual in the time interval $t$ to $t+\delta t$ is $\omega(t) S(t) \delta t+o(\delta t)$. We represent the flux of individuals entering the infected state at time $t$, by
$q^{+}(I,t)$, which can be recursively constructed from the flux at earlier times. Explicitly we have
\begin{equation}
\label{eq_flux}
q^{+}(I,t)=\int_{-\infty}^{t}\omega(t)S(t)\Phi(t,t')q^{+}(I,t')dt',
\end{equation}
where $\Phi(t,t')$ is the probability that an infected individual has survived 
in the infected state until time $t$ given that they entered the state at time 
$t'$. Let $i(-t',0)$ be the number of individuals who became infected at time $t'<0$ and who are still infected at time $0$, hence,
\begin{equation}
q^+(I,t')= \frac{i(-t',0)}{\Phi(0,t')}, \hspace{10pt}t'<0. 
\end{equation} We can then write Eq. (\ref{eq_flux}) for $t\geq 0$ as,
\begin{equation}
\label{eq_flux2}
q^{+}(I,t)=\int_{0}^{t}\omega(t)S(t)\Phi(t,t')q^{+}(I,t')dt'+\int_{-\infty}^{0}\omega(t)S(t)\frac{\Phi(t,t')}{\Phi(0,t')}i(-t',0)dt'.
\end{equation}

For an individual to be in the infected compartment at time $t$ they must have 
become infected at time $t$ or at some prior time $t'$ and remained in the compartment until 
$t$. The number infected at time $t$ can therefore be found from the flux, and the survival probability, as follows,
\begin{equation}
\label{eq_totI}
I(t)=I_0(t)+\int_{0}^{t}\Phi(t,t')q^{+}(I,t')dt'.
\end{equation}
Here we have defined the function,
\begin{equation}
I_0(t)=\int_{-\infty}^{0}\frac{\Phi(t,t')}{\Phi(0,t')}i(-t',0)dt'.
\end{equation}
We assume that there are two possible ways in which an individual can move from 
the infectious compartment; they can either recover from the disease and move 
to the $R$ compartment, or they can die and be removed from consideration. If these two 
possibilities are independent we may write,
\begin{equation}
\label{eq_sep}
\Phi(t,t')=\phi(t-t')\theta(t,t')
\end{equation}
where $\phi(t-t')$ is the probability of surviving the jump transition to the $R$ 
compartment from time $t'$ to time $t$, and $\theta(t,t')$ is the probability 
of surviving death from time $t'$ until time $t$. 
If we have a time dependent death rate such that the probability of death occurring in the interval $t$ to $t+\delta t$ is  $\gamma(t) \delta t+o(\delta t)$
then the death survival can be written as,
\begin{equation}
\theta(t,t')=e^{-\int_{t'}^{t}\gamma(s)ds}.
\end{equation}

The rate of change of the population in the infectious compartment can be found by differentiating Eq. (\ref{eq_totI}) to get,
\begin{equation}
\label{eq_didt}
\begin{split}
\frac{dI(t)}{dt}&=q^{+}(I,t)-\int_{0}^{t}\psi(t-t')\theta(t,t')q^{+}(I,
t')dt'-\gamma(t)\int_{0}^{t}\phi(t-t')\theta(t,t')q^{+}(I,t')dt'+\frac{d I_{0}(t)}{dt}\\
&=q^{+}(I,t)-\int_{0}^{t}\psi(t-t')\theta(t,t')q^{+}(I,t')dt'-\gamma(t)I(t)+\theta(t,0)\frac{d}{dt}\left(\frac{I_0(t)}{\theta(t,0)}\right).
\end{split}
\end{equation}
Here we have used the fact that, $\phi(0)=1$, and that the derivative of the jump survival function, $\phi(t)$, is a waiting time probability density function, here denoted $\psi(t)$, i.e.,
\begin{equation}
\frac{d \phi(t)}{dt}=-\psi(t).
\end{equation}
Substituting Eq. (\ref{eq_flux2}) into Eq. (\ref{eq_didt}) gives,
\begin{equation}
\label{eq_didt2}
\begin{split}
\frac{dI(t)}{dt}=&\omega(t)  
S(t)\left(\int_{0}^{t}\phi(t-t')\theta(t,t')q^{+}(I,t')dt'+I_0(t)\right)\\
&-\int_{0}^{t}\psi(t-t')\theta(t,t')q^{+}(I,t')dt'-\gamma(t)I(t)+\theta(t,0)\frac{d}{dt}\left(\frac{I_0(t)}{\theta(t,0)}\right).
\end{split}
\end{equation}
In order to obtain a generalised master equation we need to express the right 
hand side of this equation in terms of $I(t)$. We first use the definition of 
$I(t)$ in Eq. (\ref{eq_totI}), to write,
\begin{equation}
\label{eq_didt3}
\frac{dI(t)}{dt}=\omega(t)  
S(t)I(t)-\int_{0}^{t}\psi(t-t')\theta(t,t')q^{+}(I,t')dt'-\gamma(t)I(t)+\theta(t,0)\frac{d}{dt}\left(\frac{I_0(t)}{\theta(t,0)}\right).
\end{equation}
Further, noting that, 
\begin{equation}
\label{eq_semi}
\theta(t,0)=\theta(t,t')\theta(t',0)\;\;\forall\; 0<t'<t,
\end{equation}
we can write Eq. (\ref{eq_totI}) as,
\begin{equation}
\frac{I(t)}{\theta(t,0)}=\frac{I_0(t)}{\theta(t,0)}+\int_{0}^{t}\phi(t-t')\frac{q^{+}(I,t')}{\theta(t',0)}dt'.
\end{equation}
As the right hand side contains a convolution, taking the Laplace transform, $\mathcal{L}$, of the equation with respect to time and rearranging gives, 
\begin{equation}
\label{eq_ltq}
\mathcal{L}\left\{\frac{q^{+}(I,t)}{\theta(t,0)}\right\}=\mathcal{L}\left\{\frac{I(t)}{\theta(t,0)}\right\}\frac{1}{\mathcal{L}\left\{\phi(t)\right\}}-\mathcal{L}\left\{\frac{I_0(t)}{\theta(t,0)}\right\}\frac{1}{\mathcal{L}\left\{\phi(t)\right\}}.
\end{equation}
This result can then be used to write,
\begin{equation}
\label{eq_Ikap}
\begin{split}
\int_{0}^{t}\psi(t-t')\frac{q^{+}(I,t')}{\theta(t',0)}dt'&=\mathcal{L}^{-1}
\left\{\frac{\mathcal{L}\left\{\psi(t)\right\}}{\mathcal{L}\left\{\phi(t)\right\}}\mathcal{L}\left\{\frac{I(t)}{\theta(t,0)}\right\}\right\}-\mathcal{L}^{-1}
\left\{\frac{\mathcal{L}\left\{\psi(t)\right\}}{\mathcal{L}\left\{\phi(t)\right\}}\mathcal{L}\left\{\frac{I_0(t)}{\theta(t,0)}\right\}\right\},\\
&=\int_{0}^{t}K(t-t')\left(\frac{I(t')}{\theta(t',0)}-\frac{I_0(t')}{\theta(t',0)}\right)dt'.
\end{split}
\end{equation}
where we have defined the memory kernel,
\begin{equation}
\label{eq_K}
K(t)=\mathcal{L}^{-1}\left\{\frac{\mathcal{L}\left\{\psi(t)\right\}}{\mathcal
{L}\left\{\phi(t)\right\}}\right\}.
\end{equation}
Equation (\ref{eq_semi}) allows us to write Eq. (\ref{eq_didt3}) as, 
\begin{equation}
\label{eq_didt4}
\frac{dI(t)}{dt}=\omega(t)  
S(t)I(t)-\theta(t,0)\int_{0}^{t}\psi(t-t')\frac{q^{+}(I,t')}{\theta(t',0)}\, dt'-\gamma(t)I(t)+\theta(t,0)\frac{d}{dt}\left(\frac{I_0(t)}{\theta(t,0)}\right),
\end{equation}
which, using Eq. (\ref{eq_Ikap}), becomes,
\begin{equation}
\label{eq_gme}
\frac{dI(t)}{dt}=\omega(t)  
S(t)I(t)-\gamma(t)I(t)-\theta(t,0)\left(\int_{0}^{t}K(t-t')\left(\frac{I(t')}{\theta(t',0)}-\frac{I_0(t')}{\theta(t',0)}\right) dt'-\frac{d}{dt}\left(\frac{I_0(t)}{\theta(t,0)}\right)\right).
\end{equation}
This equation is the generalised master equation that describes the time 
evolution of the number of infected individuals in an SIR model with arbitrary 
waiting time in the infectious compartment. 

Simple flux balance considerations give the master equations for the other two states. The equations for the susceptible and recovered populations are,
\begin{align}
\label{eq_dSdt}
\frac{dS(t)}{dt}&=\lambda(t)-\omega(t)  S(t) I(t)-\gamma(t)S(t),\\
\label{eq_dRdt}
\frac{dR(t)}{dt}&=\theta(t,0)\left(\int_{0}^{t}K(t-t')\left(\frac{I(t')}{\theta(t',0)}-\frac{I_0(t)}{\theta(t',0)}\right) dt'-\frac{d}{dt}\left(\frac{I_0(t)}{\theta(t,0)}\right)\right)-\gamma(t)R(t).
\end{align}
Equations (\ref{eq_gme}), (\ref{eq_dSdt}), and (\ref{eq_dRdt}) are the governing, or generalised master, equations for an SIR model with a general recovery probability. 

\subsection{Relation to the Classic SIR Model}
The master equations for the SIR model with a general recovery probability reduce to the classic SIR model equations, with births and deaths, if the probability of
not clearing an infection is exponentially distributed, i.e. $\phi(t)=\exp (-\mu t)$. In this case the probability of an individual clearing an infection is not dependent on the amount of time that the person has already been infected.
Subsituting the exponential distribution for $\phi(t)$ into the kernel, Eq. (\ref{eq_K}), we obtain,
\begin{equation}
\label{eq_Kex}
K(t-t')=\mu \delta(t-t'),
\end{equation}
where $\delta(t)$ is the Dirac delta function. Also noting that as $\phi(t)=\exp (-\mu t)$ we can write,
\begin{equation}
\frac{d}{dt}\left(\frac{I_0(t)}{\theta(t,0)}\right)=-\mu\frac{I_0(t)}{\theta(t,0)}.
\end{equation}
We can now substitute the expression for the kernel, Eq. (\ref{eq_Kex}), into the generalised master equations, 
Eqs. (\ref{eq_gme}), (\ref{eq_dSdt}), and (\ref{eq_dRdt}), to yield the classic SIR equations, 
\begin{eqnarray}
\frac{dS(t)}{dt}&=&\lambda(t) -\omega(t)  S(t) I(t)-\gamma(t) S(t),\label{sir1}\\
\frac{dI(t)}{dt}&=&\omega(t)  S(t) I(t)-\mu I(t)-\gamma(t) I(t),\label{sir2}\\
\frac{dR(t)}{dt}&=&\mu I(t)-\gamma(t) R(t).\label{sir3}
\end{eqnarray}

\subsection{Relation to the Kermack and McKendrick Age-Structured Model}

The master equations for the SIR model with a general recovery probability are formally equivalent to a reduction of the general SIR model presented by Kermack and McKendrick \cite{KM1932}. The derivation of the Kermack and McKendrick model from our stochastic process is presented in Appendix A. Here we show how the master equations, Eqs. (\ref{eq_gme}), (\ref{eq_dSdt}), (\ref{eq_dRdt}), can be obtained from a reduction of the Kermack and McKendrick SIR model equations  given by, \begin{eqnarray}
\frac{d S}{dt}&=&\lambda-\int_{0}^{\infty}\omega i(a,t)da S(t)-\gamma S(t),\\
\label{eq_KMi}
\frac{\partial i}{\partial t}&+&\frac{\partial i}{\partial a}=-\beta(a) i(a,t)-\gamma i(a,t),\\
\frac{dR}{dt}&=&\int_{0}^{\infty}\beta(a) i(a,t) da-\gamma R(t),\\
\label{eq_KMI}
I(t)&=&\int_{0}^{\infty}i(a,t)da.
\end{eqnarray}
In this model $i(a,t)$ is the number of individuals who are infected at time $t$ and who have been infected since time $t-a$. 
The equivalence can be seen by making the identification,
\begin{equation}
i(a,t)=\Phi(t,t-a)q^{+}(I,t-a).
\end{equation}
Then Eq. (\ref{eq_KMI}) is equivalent to Eq. (\ref{eq_totI}), provided that the separability assumption, Eq. (\ref{eq_sep}), holds. If we assume that $i(a,t)\rightarrow 0$ as $a \rightarrow \infty$. Integrating Eq. (\ref{eq_KMi}) with respect to $a$ then gives
\begin{equation}
\frac{dI}{dt}-i(0,t)=-\int_{0}^{\infty}\beta(a)\phi(a)\theta(t,t-a)q^{+}(I,t-a)da-\gamma I.
\end{equation}
Identifying $\beta(a)\phi(a)=\psi(a)$ and $i(0,t)=q^{+}(I,t)$, we can then split the integral to,
\begin{equation}
\frac{dI}{dt}=i(0,t)-\int_0^{t} \psi(a)\theta(t,t-a)q^{+}(I,t-a)da -\int_t^{\infty} \psi(a)\theta(t,t-a)q^{+}(I,t-a)da -\gamma I.
\end{equation}
This allows for the use of the same Laplace transform method as in Eqs. (\ref{eq_ltq})- (\ref{eq_gme}), hence with a change of variable of integration $t'=t-a$ becoming,
\begin{equation}
\frac{dI}{dt}=\omega S I-\theta(t,0)\int_{0}^{t}K(t-t')\left(\frac{I(t')}{\theta(t',0)}-\frac{I_0(t')}{\theta(t',0)}\right) dt'+\theta(t,0)\frac{d}{dt}\left(\frac{I_0(t)}{\theta(t,0)}\right)-\gamma I.
\end{equation}
Using the result that,
\begin{equation}
\theta(t,0)\frac{d}{dt}\left(\frac{I_0(t)}{\theta(t,0)}\right) = -\int_{-\infty}^{0} \psi(t-t')\theta(t,t')q^{+}(I,t')dt'.
\end{equation}
Hence we have recovered the generalised master equation given in Eq. (\ref{eq_gme}) with time independent rates. 

\subsection{Integral Equation Formulation}
\label{sec_IEF}

The master equations for the fractional recovery SIR model can be formulated as a coupled set of integral equations.  This enables comparisons with other related models and it enables the application of integral equation methods for analysis of equilibrium states.
To formulate the system as integral equations  we begin by noting that Eqs.(\ref{eq_flux}), (\ref{eq_flux2}) can be substituted into Eq.(\ref{eq_totI})
to yield,
\begin{equation}
q^{+}(I,t)=\omega(t)S(t)I(t),
\end{equation}
and then Eq.(\ref{eq_totI}) can be re-written to obtain the integral equation for the time evolution of the infected state,
\begin{equation}
I(t)=I_0(t)+\int_0^t\Phi(t,t')\omega(t') S(t')I(t')\, dt'.\label{intI}
\end{equation}

The integral equation for the time evolution of the susceptible state can be obtained by direct integration of Eq. (\ref{eq_dSdt}), to yield,
\begin{equation}
S(t)=S(0)+\int_0^t\lambda(t')\, dt'-\int_0^t\omega(t')S(t')I(t')\, dt'\label{intS}
-\int_0^t\gamma(t')S(t')\, dt'.
\end{equation}

Note that the total population,
\begin{equation}
N(t)=S(t)+I(t)+R(t),\label{totN}
\end{equation}
and the master equations  with a fractional order recovery were obtained with,
\begin{equation}
\frac{dN}{dt}=\lambda(t)-\gamma(t)N(t).\label{diffN}
\end{equation}
We can combine Eqs.(\ref{totN}) and (\ref{diffN}) to obtain the differential equation for the time evolution of the recovery state in the form,
\begin{equation}
\frac{dR}{dt}=-\frac{dI}{dt}-\frac{dS}{dt}+\lambda(t)-\gamma(t)S(t)-\gamma(t)I(t)-\gamma(t)R(t),\label{dReq}
\end{equation}
and then integrate to find ,
\begin{eqnarray}
R(t)&=&R(0)-I(t)+I(0)-S(t)+S(0)+
\int_0^t \lambda(t')\, dt'\nonumber\\
& &-\int_0^t \gamma(t')S(t') \, dt'
-\int_0^t \gamma(t')I(t') \, dt'
-\int_0^t \gamma(t')R(t') \, dt'.
\end{eqnarray}
After substituting for $I(t)$ and $S(t)$ using Eqs.(\ref{intI}),(\ref{intS}) we have the integral equation for the time evolution of the recovered state, 
\begin{eqnarray}
R(t)&=&R(0)+I_0(0)-I_0(t)-\int_0^t\Phi(t,t')\omega(t') S(t')I(t')\, dt'+\int_0^t\omega(t')S(t')I(t')\, dt'\nonumber\\
& &
-\int_0^t \gamma(t')I(t') \, dt'
-\int_0^t \gamma(t')R(t') \, dt'.\label{intR}
\end{eqnarray}

Equations (\ref{intS}), (\ref{intI}), (\ref{intR}), provide a general set of coupled integral equations for fractional recovery SIR models.
The integral equation for the susceptible state, Eq. (\ref{intSc}), can be shown to be equivalent to
the integral equation obtained from $S(t)=N(t)-I(t)-R(t)$ in the special case where 
 $\lambda(t)=\gamma(t) N(t)$, and thus $N(t)=N(0)$ is constant.
\subsection{Reduction to the Hethcote and Tudor Endemic Disease Model}
\label{sec_HT}

If $\gamma(t), \omega(t)$ and $\lambda(t)$ are constant in time then the integral equations, (\ref{intS}), (\ref{intI}), (\ref{intR}) reduce to the integral equation model for endemic infection diseases that was introduced by Hethcote and Tudor \cite{HT1980}. 

We first note that, with $\gamma$ constant,
\begin{equation}
\Phi(t,t')=\phi(t-t')e^{-\gamma(t-t')},\label{Phieq}
\end{equation}
and,
\begin{equation}
\frac{dY(t)}{dt}+\gamma Y(t)=e^{-\gamma t}\frac{d}{dt}
\left(e^{\gamma t} Y(t)\right),\label{dYeq}
\end{equation}
Substituting Eq.(\ref{Phieq}) into Eq.(\ref{intI}), with $\omega$ also constant, we have,
\begin{equation}
I(t)=I_0(t)+\int_0^t\phi(t-t')e^{-\gamma(t-t')}\omega S(t')I(t')\, dt'.\label{intIc}
\end{equation}
Using Eq. (\ref{dYeq}) we can re-write Eq. (\ref{eq_dSdt}), with $\omega, \gamma$ and $\lambda$ constant as,
\begin{equation}
\frac{d}{dt}
\left(e^{\gamma t} S(t)\right)
=e^{\gamma t} \lambda -e^{\gamma t}\omega S(t)I(t),
\end{equation}
and then integrate with respect to time to obtain,
\begin{equation}
S(t)=e^{-\gamma t}S(0)+
\int_0^t e^{-\gamma(t-t')}\lambda \, dt'
-\int_0^te^{-\gamma(t-t')}\omega S(t')I(t')\, dt'.\label{intSc}
\end{equation}
Using Eq. (\ref{dYeq}) we can re-write Eq. (\ref{dReq}), with $\omega, \gamma$ and $\lambda$ constant as,
\begin{equation}
\frac{d}{dt}\left(e^{\gamma t}R(t)\right)=\lambda e^{\gamma t}
-\frac{d}{dt}\left(e^{\gamma t}S(t)\right)
-\frac{d}{dt}\left(e^{\gamma t}I(t)\right),
\end{equation}
and then integrate with respect to time to obtain,
\begin{equation}
R(t)=R(0)e^{-\gamma t}
-I(t)+I(0)e^{-\gamma t}-S(t)+S(0)e^{-\gamma t}
+\int_0^t \lambda e^{-\gamma(t-t')}\, dt'.
\end{equation}
We now substitute for $I(t)$ and $S(t)$ using Eqs.(\ref{intIc}), (\ref{intSc}) to obtain,
\begin{equation}
R(t)=R(0)e^{-\gamma t}+I_0(0)e^{-\gamma t}-I_0(t)
+\int_0^t\omega S(t')I(t')e^{-\gamma(t-t')}\left(1-\phi(t-t')\right)\, dt'.\label{intRc}
\end{equation}
Equations (\ref{intIc}) and (\ref{intRc}) recover the integral equations for the infected state and the recovery state in the endemic infection diseases model introduced
in  \cite{HT1980}. 
The integral equation for the susceptible state in this case
 can be shown to be equivalent to
the integral equation obtained from $S(t)=N-I(t)-R(t)$ with
 $N(t)=N(0)=\lambda/\gamma$.

\section{Fractional Recovery SIR Model}
\label{sec_fSIR}
When a person is persistently infected for a long period period of time, with little chance of spontaneous recovery, they are said to be chronically infected. This type of behaviour is not captured by the assumptions of the standard SIR model, i.e., exponentially distributed waiting times. We can incorporate chronic infections by having the probability of clearing the disease decrease with the amount of time that an individual has been infected. In a power-law tailed waiting time distribution the expected waiting time diverges. Using such a distribution in our SIR model will lead to individuals becoming ``trapped" in the infectious compartment until they die. By utilising a power-law tailed Mittag-Leffler waiting time distribution our general SIR model will reduce to a set of differential equations with a fractional order time derivative on the recovery transition.   

The Mittag-Leffler probability density is defined by \cite{HA1995},
 \begin{equation}
 \psi(t)=\frac{t^{\alpha -1}}{\tau^\alpha} E_{\alpha,\alpha}\left(-\left(\frac{t}{\tau}\right)^\alpha\right),
 \end{equation}
 with $0<\alpha\leq1$. Here $E_{\alpha,\beta}(z)$ is the two parameter Mittag-Leffler function, defined by,
 \begin{equation}
 E_{\alpha,\beta}(z)=\sum_{k=0}^\infty\frac{z^k}{\Gamma(k\alpha+\beta)}.
 \end{equation}
 The Mittag-Leffler distribution limits to an exponential distribution in the case $\alpha=1$.
 For $0<\alpha<1$ the density has a power-law tail at long times \cite{BE1953}, 
 \begin{equation}
 \psi(t)\sim t^{-1-\alpha}.
 \end{equation}
 The corresponding survival function is given by,
 \begin{equation}
 \label{eq_surv}
\phi(t)=E_{\alpha,1}\left(-\left(\frac{t}{\tau}\right)^\alpha\right).
\end{equation}

The Laplace transform from $t$ to $s$ of the memory kernel, Eq. (\ref{eq_K}), for a Mittag-Leffler distribution is given by, 
\begin{equation}
\label{eq_MLK}
\mathcal{L}_t [K(t)| s]=s^{1-\alpha}\tau^\alpha.
\end{equation}
Here we use the notation $\mathcal{L}_t [Y(t)| s]$ to denote the Laplace transform
of $Y(t)$ from $t$ to $s$ and we use the notation $\mathcal{L}_s^{-1} [Y(s)| t]$
to denote the inverse Laplace transform
of $Y(s)$ from $s$ to $t$.
Without loss of generality we define,
\begin{equation}
\mu=\tau^\alpha.
\end{equation}
We also note that the Riemann-Louiville fractional derivative,  defined by \cite{OS1974}
\begin{equation}
\,_0\mathcal{D}_{t}^{1-\alpha}f(t)=\frac{1}{\Gamma(\alpha)}\frac{d}{dt}\int_{0}^t \frac{f(t')}{(t-t')^{1-\alpha}}\, dt'
\end{equation}
has the Laplace transform \cite{P1999}
 \begin{equation}
\mathcal{L}_t[\,_0\mathcal{D}_{t}^{1-\alpha}f(t)\vert s]=s^{1-\alpha}
\mathcal{L}_t[f(t)|s]
\end{equation}
so that the fractional derivative can also be written
 as the following convolution,
\begin{equation}
\,_0\mathcal{D}_{t}^{1-\alpha}f(t)=\int_{0}^{t}\mathcal{L}_s^{-1}\left[s^{1-\alpha}\big\vert t'
\right] f(t-t')dt'.\label{convolve}
\end{equation}
It follows from Eq. (\ref{eq_MLK}) and Eq. (\ref{convolve}) that if the kernel in  Eqs. (\ref{eq_gme}), (\ref{eq_dSdt}), (\ref{eq_dRdt}) is obtained from the Mittag-Leffler waiting time density then we obtain
 the 
generalised master equations with fractional recovery,
\begin{align}
\label{eq_fS}
\frac{dS(t)}{dt}&=\lambda(t) -\omega(t)  S(t) I(t)-\gamma(t) S(t),\\
\label{eq_fI}
\frac{dI(t)}{dt}&=\omega(t)  S(t) I(t)-\gamma(t) I(t)-\theta(t,0)\left(\mu\mathcal{D}_{t}^{1-\alpha}\left(\frac{I(t)}{\theta(t,0)}-\frac{I_0(t)}{\theta(t,0)}\right)-\frac{d}{dt}\left(\frac{I_0(t)}{\theta(t,0)}\right)\right),\\
\label{eq_fR}
\frac{dR(t)}{dt}&=\theta(t,0)\left(\mu\mathcal{D}_{t}^{1-\alpha}\left(\frac{I(t)}{\theta(t,0)}-\frac{I_0(t)}{\theta(t,0)}\right)-\frac{d}{dt}\left(\frac{I_0(t)}{\theta(t,0)}\right)\right)-\gamma(t) R(t).
\end{align}
In the following we will refer to the above set of equations as the frSIR model. Letting $\alpha=1$ recovers the standard SIR model.

\subsection{Equilibrium States}

The frSIR model is a non-autonomous dynamical system which can be simplified, by taking the birth, infectivity and death rate to be constant parameters, $\lambda(t)=\lambda$, $\omega(t)=\omega$ and $\gamma(t)=\gamma$ respectively, giving,
\begin{align}
\frac{dS(t)}{dt}&=\lambda -\omega S(t) I(t)-\gamma S(t),\label{fsir1}\\
\frac{dI(t)}{dt}&=\omega  S(t) I(t)-\gamma I(t)-e^{-\gamma t}\left(\mu\,_0\mathcal{D}_{t}^{1-\alpha}\left(e^{\gamma t} \left(I(t)-I_0(t)\right)
\right)-\frac{d}{dt}\left(e^{\gamma t}I_{0}(t)\right)\right),\label{fsir2}\\
\frac{dR(t)}{dt}&=e^{-\gamma t}\left(\mu\,_0\mathcal{D}_{t}^{1-\alpha}\left(e^{\gamma t} \left(I(t)-I_0(t)\right)
\right)-\frac{d}{dt}\left(e^{\gamma t}I_{0}(t)\right)\right)-\gamma R(t).\label{fsir3}
\end{align}
We take the equilibrium state to be ($S^*, I^*, R^*$), such that,
\begin{equation}
\lim_{t \rightarrow \infty}S(t)=S^*,\hspace{25pt}\lim_{t \rightarrow \infty}I(t)=I^*,\hspace{25pt}\lim_{t \rightarrow \infty}R(t)=R^*.
\end{equation}
Taking the limit as $t \rightarrow \infty$ of Eqs. (\ref{fsir1}), (\ref{fsir2}), and (\ref{fsir3}), and noting that,
\begin{equation}
\lim_{t\rightarrow \infty} I_0(t)=0,\,\,\lim_{t\rightarrow \infty} \frac{d}{dt}\left(e^{\gamma t} I_0(t)\right) = 0.
\end{equation}
we have
\begin{align}
0&=\lambda -\omega S^* I^*-\gamma S^*,\\
0&=\omega S^* I^*-\lim_{t \rightarrow \infty} e^{-\gamma t}\mu\,_0\mathcal{D}_{t}^{1-\alpha}\left(e^{\gamma t}\left(I(t)-I_0(t)\right)\right), \label{fsir2a}\\
0&=\lim_{t \rightarrow \infty} e^{-\gamma t}\mu\,_0\mathcal{D}_{t}^{1-\alpha}\left(e^{\gamma t}\left(I(t)-I_0(t)\right)\right)-\gamma R^*. \label{fsir3a}
\end{align}

In order to calculate the unevaluated limits in Eqs.  (\ref{fsir2a}), and (\ref{fsir3a})  we consider a Laplace transform of the terms,
\begin{equation}
\label{eq_Laplace}
\mathcal{L}\{e^{-\gamma t} \,_0\mathcal{D}_{t}^{1-\alpha} \left(e^{\gamma t} J(t)\right)\}=(s+\gamma)^{1-\alpha}\left(\hat{J}(s)\right).
\end{equation}
In which we have defined,
\begin{equation}
J(t)=I(t)-I_0(t).
\end{equation}
We can express Eq. (\ref{eq_Laplace}) using a Taylor series expansion,
\begin{equation}
\label{eq_taylorLaplace}
\hat{J}(s)(s+\gamma)^{1-\alpha}=\hat{J}(s)\left(\gamma^{1-\alpha}+(1-\alpha)
\gamma^{-\alpha}s +O(s^2)\right).
\end{equation}
As the Laplace transform is a linear operator we can take the inverse termwise, producing,
\begin{align}
e^{-\gamma t} \,_0\mathcal{D}_{t}^{1-\alpha} \left(e^{\gamma t} (I(t)-I_0(t))\right)&=\mathcal{L}^{-1}\{\hat{J}(s)\left(\gamma^{1-\alpha}+(1-\alpha)
\gamma^{-\alpha}s +O(s^2)\right)\},\\
&=\gamma^{1-\alpha}J(t)+(1-\alpha)\gamma^{-\alpha}\frac{dJ(t)}{dt}+\mathcal{L}^{-1}\left(O(s^2)\right).
\end{align}
The limit of $J(t)$ is,
\begin{equation}
\lim_{t \rightarrow \infty}J(t)=I^*.
\end{equation}
It is then clear that, in the long time limit, the inverse Laplace transform of the higher order terms of the Taylor expansion will become zero, i.e.
\begin{align}
\lim_{t\to0}\frac{dJ(t)}{dt}&=0,\\
\lim_{t\to0}\mathcal{L}^{-1}\left(O(s^2)\right)&=0.
\end{align}
Thus we can compute the desired limit,
\begin{equation}
\label{longD}
\lim_{t\to\infty} e^{-\gamma t}\,_0\mathcal{D}_{t}^{1-\alpha}\left(e^{\gamma t} \left(I(t)-I_0(t)\right)\right)=\gamma^{1-\alpha}I^*.
\end{equation}
Substituting Eq. (\ref{longD}) into Eqs. (\ref{fsir1}), (\ref{fsir2}), and (\ref{fsir3}) yields,
\begin{align}
\label{fsir1st}
0 &= \lambda -\omega S^* I^* -\gamma S^*,\\
\label{fsir2st}
0 &= \omega S^* I^* -\mu \gamma^{1-\alpha}I^* -\gamma I^*,\\
\label{fsir3st}
0 &= \mu\gamma^{1-\alpha}I^* - \gamma R^*.
\end{align}
Solving Eqs. (\ref{fsir1st}), (\ref{fsir2st}), and (\ref{fsir3st}) reveals two equilibrium states, the disease free state,
\begin{equation} \label{eq_ss1}
S^*=\frac{\lambda}{\gamma},\;\;\;I^*=0,\;\;\;R^*=0,
\end{equation}
and the endemic state,
\begin{equation}
\label{eq_ss2}
S^{*}=\frac{\mu\gamma^{1-\alpha}+\gamma}{\omega},\;\;\;I^{*}=\frac{\lambda}{\mu\gamma^{1-\alpha}+\gamma}-\frac{\gamma}{\omega},\;\;\;R^{*}=\frac{\mu\lambda}{\mu\gamma+\gamma^{1+\alpha}}-\frac{\mu\gamma^{1-\alpha}}{\omega}.
\end{equation}
As the population in each compartment can not be negative the endemic equilibrium can only exist if,
\begin{equation}
\label{eq_cond1}
\lambda\omega>\mu\gamma^{2-\alpha}+\gamma^2.
\end{equation}

When $\alpha=1$,  the equilibrium states are equivalent to the fixed points of the classic SIR model with constant parameters. 
For $0<\alpha<1$ the equilibrium states are not fixed points. This invalidates the use of a standard linear stability analysis
around the equilibrium states. However some progress can be made by considering the integral equation formulation of the frSIR model, Eqs. (\ref{intIc}), (\ref{intSc}), and (\ref{intRc}). After a translation of the equilibrium states to the origin, the asymptotic behaviour of the resulting system of nonlinear Volterra integral equations is equivalently given  by the asymptotic
behaviour of its  linearization as a  system of linear Volterra integral equations \cite{M1968}. This remarkable result was
used by Hethcote and Tudor \cite{HT1980} to infer local stability properties of the equilibrium states of the integral equations  Eqs. (\ref{intIc}), (\ref{intSc}), and (\ref{intRc}). If the results of Hethcote and Tudor \cite{HT1980} are applied to
 the special case of the frSIR model, where  $\phi(t)$ is defined in Eq.(\ref{eq_surv}), then the disease free state, Eq.(\ref{eq_ss1}), is locally stable if $\lambda\omega\le\mu\gamma^{2-\alpha}+\gamma^2$ and the endemic equilibrium state, Eq.(\ref{eq_ss2}), is locally stable when it exists.

\section{SIR as a Discrete Time Random Walk}
\label{sec_numerical}
There are numerous numerical methods that have been developed for solving fractional order differential equations
\cite{OS1974,P1999,DFF2002} and many of these methods could be adapted to the system under consideration here.
However recently we showed that in the case where the fractional order derivatives
have been derived from continuous time random walks it is useful to reformulate the problem using discrete time random walks (DTRWs) and then use this formulation as the basis of a numerical method \cite{ADHN2014}.
The advantage of basing the numerical method on a discrete time stochastic process is that the derived explicit numerical method is easy to implement and is also inherently stable.
\subsection{Discrete Time Random Walk}

We consider a discrete time random walk where the walking particle is an individual who will transition though the S, I, and R, compartments. 
In order to obtain a useful numerical scheme from this we need the discrete time process to limit to the continuum process as the time step, $\Delta t$, goes to zero. This necessitates a slight modification to the transitions that were considered in the continuum case. Individuals are born into the susceptible compartment with a birth rate $\lambda(t)$ so that the number of individuals being born on the $n^{\mathrm{th}}$ time step, between time $t$ and $t+\Delta t$, is equal to $\Lambda(n)=\lambda(t) \Delta t$. The probability that an individual will die and be removed from consideration on the $n^{\mathrm{th}}$ time step is $\gamma(n) $. Susceptible individuals who come in contact with an infected individual may become infected. The probability of an infected individual coming in contact and infecting a susceptible individual in the $n^{\mathrm{th}}$ time step is $\omega(n)$.

The recovery of infected individuals is assumed to be dependent on the number of time steps since they entered the infectious compartment. The individual is assumed to wait in the infectious compartment with a probability of ``jumping'' that is dependent on the time step. When a transition event occurs there are two possible things that can happen to the individual, they will either move to the recovered compartment or re-enter the infectious compartment with the transition probability, dependent on the time step, being reset. If the individual was infected before the first time step they will always transition to the recovered compartment. This self jump modification is required to ensure appropriate scaling as the size of the time step tends to zero. The probability of transitioning to the $R$ compartment, given a jump occurred, is denoted by $r$.  The probability of jumping on the $n^\mathrm{th}$ step, conditional on not having jumped in the first $(n-1)$ steps, is denoted by $\mu(n)$. 

From this it follows that the probability flux entering the $I$ compartment on the $n^\mathrm{th}$ time step can be written as
\begin{equation}
\label{eq_dQ0}
\begin{split}
Q_I^+(n)=&\sum_{k=-\infty}^{n-1}\omega(n) S(n-1)\phi(n-1-k)\theta(n-1,k)Q_I^+(k)\\
&+(1-r)\sum_{k=1}^{n-1}\mu(n-k) \phi(n-1-k)\theta(n,k)Q_I^+(k).
\end{split}
\end{equation} 
In the above equation, $\phi(n-k)$ is the probability of not jumping for $(n-k)$ steps and $\theta(n,k)$ is the probability that an individual who entered the $I$ compartment on the  $k^\mathrm{th}$ step has survived the death process
up to the $n^\mathrm{th}$ step. If we assume that we have an initial distribution of infectious individuals at the $0^{\mathrm{th}}$ time step, then we can infer that the flux at earlier times is given by,
\begin{equation}
Q_I^+(n)=\frac{i(-n,0)}{\theta(0,-n)\phi(-n)}\;\;\forall n\leq 0.
\end{equation}
Here $i(n,0)$ is the number of individuals at time step $0$ who have been infected since time step $n$, with $n\leq 0$. This allows us to write the flux for $n>0$ as 
\begin{equation}
\label{eq_dQ}
\begin{split}
Q_I^+(n)=&\omega(n) S(n-1)\sum_{k=1}^{n-1}\phi(n-1-k)\theta(n-1,k)Q_I^+(k)\\
&+(1-r)\sum_{k=1}^{n-1}\mu(n-k) \phi(n-1-k)\theta(n,k)Q_I^+(k)\\
&+\omega(n) S(n-1)\theta(n-1,0)\sum_{k=-\infty}^{0}\phi(n-1-k)\frac{i(-k,0)}{\phi(-k)}.\\
\end{split}
\end{equation}

We can easily see that 
\begin{equation}
\phi(n-k)=\prod_{j=0}^{n-k}(1-\mu(j)),
\end{equation}
and 
\begin{equation}
\theta(n,k)=\prod_{j=k}^{n}(1-\gamma(j)).
\end{equation}

The number of individuals in the $I$ compartment on the $n^{\mathrm{th}}$ time step is
\begin{equation}
\label{eqIn}
I(n)=\sum_{k=1}^{n}\phi(n-k)\theta(n,k)Q^{+}_{I}(k)+I_0(n).
\end{equation}
where,
\begin{equation}\label{eq_DI0}
I_0(n)=\theta(n,0)\sum_{k=-\infty}^{0}\phi(n-k)\frac{i(-k,0)}{\phi(-k)}.
\end{equation}
Subtracting $I(n-1)$ from each side of Eq. (\ref{eqIn}) gives,
\begin{equation}
\label{eq_dP}
\begin{split}
I(n)-I(n-1)&=Q^+_{I}(n)+\sum_{k=1}^{n-1}(\phi(n-k)\theta(n,k)-\phi(n-1-k)\theta(n-1,k))Q^+_I(k)\\
&+I_0(n)-I_0(n-1),
\end{split}
\end{equation}
and substituting Eq. (\ref{eq_dQ}) into the right hand side of Eq. (\ref{eq_dP}) and using Eq.(\ref{eqIn}) gives,
\begin{equation}
\begin{split}
I(n)-I(n-1)=&\omega(n) S(n-1)I(n-1))+I_0(n)-I_0(n-1)\\
&+(1-r)\sum_{k=1}^{n-1}\mu(n-k) \phi(n-1-k)\theta(n,k)Q_I^+(k)\\
&+\sum_{k=1}^{n-1}(\phi(n-k)\theta(n,k)-\phi(n-1-k)\theta(n-1,k))Q^+_I(k)\\
\end{split}
\end{equation}
Noting that,
\begin{equation}
(\phi(n-k)\theta(n,k)-\phi(n-1-k)\theta(n-1,k))=-\gamma(n) \phi(n-1-k)\theta(n-1,k)-\mu(n-k)\phi(n-1-k)\theta(n,k).
\end{equation}
we get,
\begin{equation}
\begin{split}
I(n)-I(n-1)=&\omega(n) S(n-1)I(n-1)+I_0(n)-I_0(n-1)-\gamma(n) (I(n-1)-I_0(n-1))\\
&-r\sum_{k=1}^{n-1}\mu(n-k)\phi(n-1-k)\theta(n,k)Q^+_I(k).\label{Iabove2}
\end{split}
\end{equation}

To simplify further we mirror the derivation in the CTRW approach, using the semi-group property
of the death survival function, and using discrete $\mathcal{Z}$ transform methods to replace the memory kernels.
The $\mathcal{Z}$ transform form $n$ to $z$ of $Y(n)$ is defined by
\begin{equation}
\mathcal{Z}[Y(n)|z]=\sum_{n=0}^\infty Y(n) z^{-n}.
\end{equation}
First we use the result
\begin{equation}
\theta(n,0)=\theta(n,k)\theta(k,0),
\end{equation}
in Eq.(\ref{Iabove2}) to write
\begin{equation}
\label{eq_dme}
\begin{split}
I(n)-I(n-1)=&\omega(n) S(n-1)I(n-1)+I_0(n)-I_0(n-1)-\gamma(n) (I(n-1)-I_0(n-1))\\
&-r \theta(n,0)\sum_{k=1}^{n-1}\mu(n-k)\phi(n-1-k)\frac{Q^+_I(k)}{\theta(k,0)}.
\end{split}
\end{equation}
and we use the same result in Eq. (\ref{eqIn}) to write 
\begin{equation}
\label{eqInm1}
\frac{I(n)-I_0(n)}{\theta(n,0)}=\sum_{k=1}^{n}\phi(n-k)\frac{Q^{+}_{I}(k)}{\theta(k,0)}.
\end{equation}
Finally, we need to express the sum $\sum_{k=1}^{n-1}\mu(n-k)\phi(n-1-k)\frac{Q^+_I(k)}{\theta(k,0)}$, in terms of $I(n)$ so we take the Z-transform of Eq. (\ref{eqInm1}), and use the convolution property of $\mathcal{Z}$ transforms to give,
\begin{equation}
\label{eqZInm1}
\mathcal{Z}\left[\frac{I(n)-I_0(n)}{\theta(n,0)}|z\right]=\mathcal{Z}\left[\phi(n)\right|z]\mathcal{Z}\left[\frac{Q^{+}_{I}(n)}{\theta(n,0)}|z\right].
\end{equation}
We can now write
\begin{equation}
\label{eq_mk}
\sum_{k=0}^{n-1}\mu(n-k)\phi(n-1-k)\frac{Q^+_I(k)}{\theta(k,0)}=\sum_{k=0}^{n-1}\kappa(n-1-k)\frac{I(k)-I_0(k)}{\theta(k,0)}
\end{equation}
where,
\begin{equation}
\kappa(n)=\mathcal{Z}^{-1}
\left[
\frac{\mathcal{Z}\left[\mu(n)\phi(n-1)|z\right]}{\mathcal{Z}\left[\phi(n)|z\right]}
\bigg\vert n\right].
\label{kappa}
\end{equation}
The discrete time generalised master equation for an arbitrary waiting time distribution can then finally be found by substituting 
Eq. (\ref{eq_mk}) into Eq. (\ref{eq_dme}) to give,
\begin{equation}
\label{eq_DI}
\begin{split}
I(n)-I(n-1)=&\omega(n)  S(n-1)I(n-1)+I_0(n)-I_0(n-1)-\gamma(n) (I(n-1)-I_0(n-1))\\
&-r\theta(n,0)\sum_{k=1}^{n-1}\kappa(n-k)\frac{I(k)-I_0(k)}{\theta(k,0)}.
\end{split}
\end{equation}
The master equations for the other compartments can be found by again considering a flux balance so that,
\begin{align}
\label{eq_DS}
S(n)-S(n-1)&=\Lambda(n)-\omega(n)S(n-1)I(n-1)-\gamma(n)S(n-1),\\
\label{eq_DR}
R(n)-R(n-1)&=r\theta(n,0)\sum_{k=1}^{n-1}\kappa(n-k)\frac{I(k)-I_0(k)}{\theta(k,0)}\\
&+I_0(n)-I_0(n-1)+\gamma(n)I_0(n-1)-\gamma(n) R(n-1).\nonumber
\end{align}

\subsection{Discrete Time Fractional Recovery SIR Model}
The continuous time frSIR model was obtained by considering Mittag-Leffler waiting time densities in the CTRW
formulation. In the DTRW formulation a discrete time fractional recovery SIR model is obtained by
considering a Sibuya($\alpha$) waiting time distribution \cite{S1979,ADHN2014}. In this case the probability of jumping on the $n^{\mathrm{th}}$ time step, conditional on not having jumped on the previous $n-1$ time steps, is given by 
\begin{equation}
\mu(n)=\left\{\begin{array}{ll}
0,&\quad n=0,\cr
\frac{\alpha}{n},&\quad n>0,\cr
\end{array}\right.
\end{equation}
and the jump survival probability is given by
\begin{equation}
\phi(n)=\frac{\Gamma(1-\alpha+n)}{\Gamma(n+1)\Gamma(1-\alpha)}.
\end{equation}
It is a simple matter to obtain the $\mathcal{Z}$ transforms from $n$ to $z$
\begin{equation}
\mathcal{Z}\left[\phi(n)|z\right]=\left(\frac{z-1}{z}\right)^{\alpha-1}
\end{equation}
and 
\begin{eqnarray}
\mathcal{Z}\left[\mu(n)\phi(n-1)|z\right]&=&\sum_{n=0}^{\infty}\mu(n)\phi(n-1)z^{-n}\\
&=&\sum_{n=1}^{\infty}\mu(n)\phi(n-1)z^{-n}\\
&=&1-\left(\frac{z-1}{z}\right)^{\alpha}.
\end{eqnarray}
Now using Eq.(\ref{kappa}) we have
\begin{equation}
\kappa(n)=\mathcal{Z}^{-1}\left[(1-z^{-1})^{1-\alpha}-(1-z^{-1})\vert n\right]\label{Zkappa}
\end{equation}
and then after taking the inverse Z transform we obtain the fractional memory kernel
\begin{equation}
\kappa(n)=\delta_{1,n}+\frac{\Gamma(n-1+\alpha)}{\Gamma(\alpha-1)\Gamma(n+1)}.\label{fkappa}
\end{equation}
This memory kernel can be calculated recursively by noting that for $n\geq3$, 
\begin{equation}
\kappa(n)=\left(1+\frac{\alpha-1}{n}\right)\kappa(n-1).
\end{equation} 
The first two values are simply $\kappa(2)=\frac{\alpha}{2}(\alpha-1)$and $\kappa(1)=\alpha$. 

The number of initially infected individuals left at the $n^{\mathrm{th}}$time step, $I_0(n)$, is found from Eq. (\ref{eq_DI0}) for the case $0<\alpha<1$,
\begin{equation}
I_0(n)=\theta(n,0)\sum_{k=-\infty}^{0} \frac{\Gamma(1-k)\Gamma(1-k+n-\alpha)}{\Gamma(1-k+n)\Gamma(1-k-\alpha)}i(-k,0).\label{eq_I0}
\end{equation}
In the case where the initially infected individuals were all infected at time zero this will simplify. Taking $i(-k,0)=i_0 \delta_{k,0}$ where $i_0$ is a constant and $\delta_{k,0}$ is a Kronecker Delta function, we find,
\begin{equation}
I_0(n)=\theta(n,0) \frac{\Gamma(1+n-\alpha)}{\Gamma(1+n)\Gamma(1-\alpha)}i_0.
\end{equation}
Again this may be also expressed recursively,
\begin{equation}
I_0(n)=\left(1-\frac{\alpha}{n}\right)I_0(n-1),
\end{equation}
with the initial condition, $I_0(0)=i_0$. 

When $\alpha=1$, $\phi(n)=0$, for all $n>0$, as such care has to be taken with the definition of $I_0(n)$. In this case the only physically permitted initial condition is that the initially infected individuals were all infected at time zero. Hence, when $\alpha=1$, $I_0(n)=0$ for $n>0$, and $I_0(0)=i_0$.  

The discrete time fractional recovery SIR model is obtained by using the memory kernel, Eq. (\ref{fkappa}), in Eqs.(\ref{eq_DI}), (\ref{eq_DS}), and (\ref{eq_DR}).

In Appendix B we show that the discrete time fractional recovery SIR model equations limit to the frSIR model equations by identifying $t=n\Delta t$ and taking the limit $\Delta t\to 0$ and $r \to 0$, with 
\begin{equation}
\lim_{\Delta t,r\to 0}\frac{r}{\Delta t^\alpha}=\mu.\label{limrt}
\end{equation}
This justifies our use of the explicit discrete time model equations as a numerical method for solving the continuous time frSIR model.

\subsection{Equivalence Between The Discrete and Continuous Parameters}
As the discrete frSIR governing equations, Eqs. (\ref{eq_DS}),  (\ref{eq_DI}), and (\ref{eq_DR}) with Eq. (\ref{fkappa}), limit to the continuous time frSIR model equations we can approximate the solution of the continuous time equations with the solution to the discrete time equations. Given a set of parameters for the continuous time equations we need to identify corresponding values of the parameters in the discrete time equations. The continuous time frSIR model is parameterised by three functions, ($\lambda(t)$, $ \omega(t)$, $\gamma(t)$), and two constants ($\alpha$, $\mu$). The discrete time frSIR model is parameterised by three functions ($\Lambda(n)$, $\omega(n)$, $\gamma(n)$), and three constants ($\alpha$, $r$, $\Delta t$). Assuming a given time step $\Delta t$, the correspondence between the continuous time $t$ and the discrete time $n$ is given by $t=n\Delta t$. 
The expected number of births in a time step is related to the continuous time birth rate by,
\begin{equation}
\Lambda(n)= \Delta t \lambda(n\delta t).
\end{equation} 
The probability of an individual becoming infected in a time step is related to the continuous time infection rate by,
\begin{equation}
\omega(n)=1- \exp\left(-\int_{n \delta t}^{(n+1)\Delta t}\omega(t')\, dt'\right).\label{omegan}
\end{equation}
Note that in the limit $\Delta t\to 0$ Eq.(\ref{omegan}) is consistent with Eq.(\ref{omegat}).
We can treat the death probability in a similar fashion, 
\begin{equation}
\gamma(n)=1- \exp\left(-\int_{n \delta t}^{(n+1)\Delta t}\gamma(t')\, dt'\right).\label{gamman}
\end{equation}
This definition of the death probability ensures that the discrete survival function, $\theta(n,m)$ is equal to the continuous time survival function evaluated at $n\delta t$ $m\delta t$, i.e.
\begin{equation}
\theta(n,m)=\exp\left(-\int_{m \Delta t}^{n\Delta t}\gamma(t')\, dt'\right).
\end{equation}
The discrete anomalous exponent, $\alpha$, is unchanged from the continuous case. 
The remaining parameter in the discrete model, $r$, is obtained from
\begin{equation}
r=\mu \Delta t^\alpha\label{req}
\end{equation}
which is consistent with Eq.(\ref{limrt}).
Note that $r$ is a probability and as such is bound in the interval $[0,1]$. Thus given the parameters
$\mu$ and $\alpha$ from the continuous time model,  Eq.(\ref{req}) constrains $\Delta t$ as follows,
\begin{equation}
\Delta t<\left(\frac{1}{\mu}\right)^{\frac{1}{\alpha}}.
\end{equation}
Given that Eqs. (\ref{eq_DS}),  (\ref{eq_DI}), and (\ref{eq_DR}) are explicit difference equations if we know $S(k)$, $I(k)$, and $R(k)$, for $1<k<n$, then we can calculate $S(n)$, $I(n)$, and $R(n)$. In this manner we have a simple numerical method that approximates the solution of the continuous time frSIR model by simply noting that for $t=n\Delta t$,
$S(t) \approx S(n)$, $I(t) \approx I(n)$, and $R(t) \approx R(n)$.

\section{Example}
\label{sec_example}
\begin{figure}
\begin{center}
\includegraphics[width=100mm]{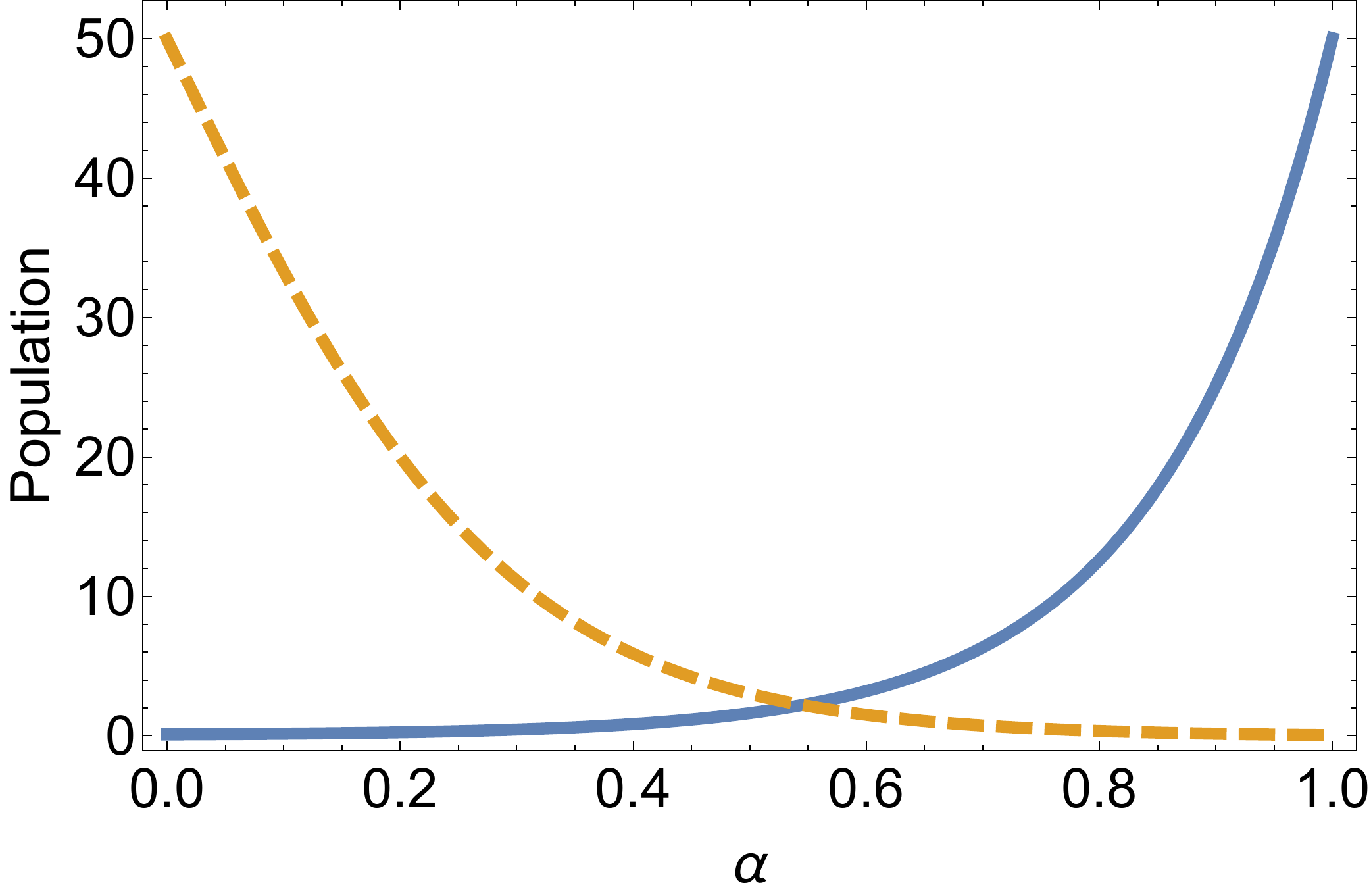}
\caption{The endemic steady state in the frSIR model plotted as a function of $\alpha$ for susceptibles (solid line) and infecteds (dashed line). The Parameters were $\lambda=0.1$, $\omega=0.02$, $\gamma=0.001$.}
\label{fig_SS}
\end{center}
\end{figure}

We consider the frSIR model, Eqs. (\ref{eq_fS}), (\ref{eq_fI}), and (\ref{eq_fR}), with the following parameters, $\lambda(t)=0.1$, $\omega(t)=0.02$, $\gamma(t)=0.001$, $\mu=1$, and a range of $\alpha\in(0,1]$. We also take a delta function initial condition at $t=0$ for the infected population, $i(t,0)=0.5\delta(t)$. The initial recovered population is taken to be zero, $R(0)=0$, and the initial population in the susceptible compartment is taken so that the total population is at the equilibrium level of $100$, i.e. $S(0)=99.5$.   
With these parameter values, the system has two possible steady states, 
the disease free steady state given by Eq. (\ref{eq_ss1}) and the endemic steady state given by Eq. (\ref{eq_ss2}).  
The endemic steady state for susceptibles and infecteds are plotted as a function of $\alpha$ in Fig.(\ref{fig_SS}).

To find the numerical approximation for this situation we solve the discrete equations, Eqs.(\ref{eq_DS}), (\ref{eq_DI}), and (\ref{eq_DR}). For a given $\Delta t$, the discrete parameters are taken to be $\Lambda(n)=0.1\Delta t$, $\omega(n)=1-\exp(-0.02 \Delta t)$, $\gamma(n)=1-\exp(-0.001 \Delta t)$, and $r=\Delta t^{\alpha}$. The initial conditions are implemented by taking $S(0)=99.5$, $R(0)=0$, and $i(-k,0)=0.5\delta_{0,k}$.

\begin{figure}
\begin{center}
\includegraphics[width=\textwidth]{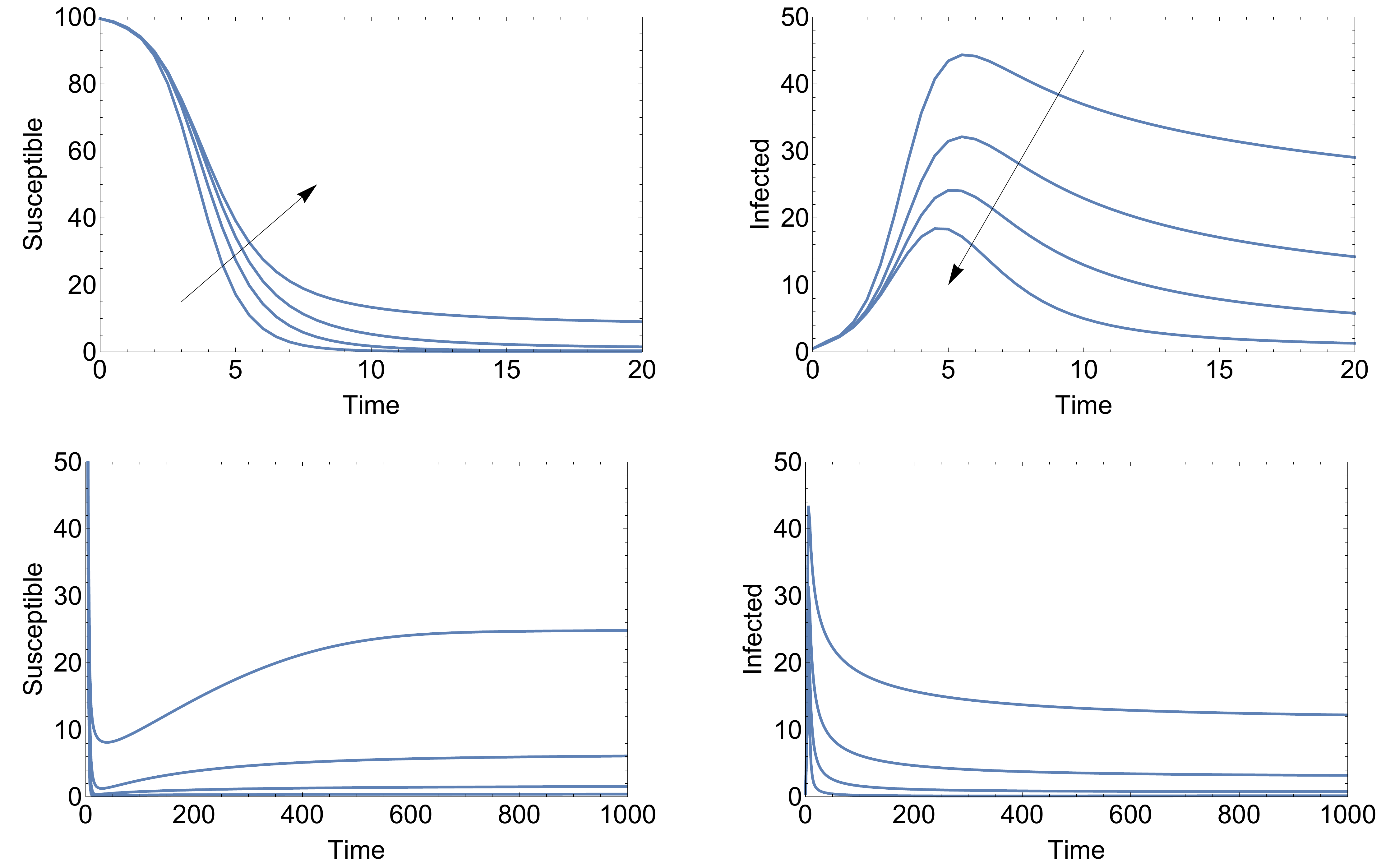}
\caption{Plots of $S(t)$ and $I(t)$ versus time in the fractional recovery SIR model with $\alpha=0.3, 0.5, 0.7, 0.9$. The arrow indicates the direction of increasing $\alpha$.
The other parameters are $\lambda=0.1$, $\omega=0.02$, $\gamma=0.001$. The model was solved using the DTRW method with $\Delta t=0.05$. }
\label{fig2}
\end{center}
\end{figure}

In Fig. (\ref{fig2}) we show plots of $S(t)$ and $I(t)$ versus time for $\alpha=0.3, 0.5, 0.7, 0.9$ with initial conditions
$S(0)=99.5, I(0)=0.5, R(0)=0$.
The number of susceptibles falls sharply over short times before rising slowly towards a steady state at later times. The number of infecteds rises to a maximum in the short time regime before
declining slowly back towards a long time steady state. It can be seen that the peak level of infection and the long time levels of infection are both increased with decreasing $\alpha$. 

\begin{figure}
\begin{center}
\includegraphics[width=100mm]{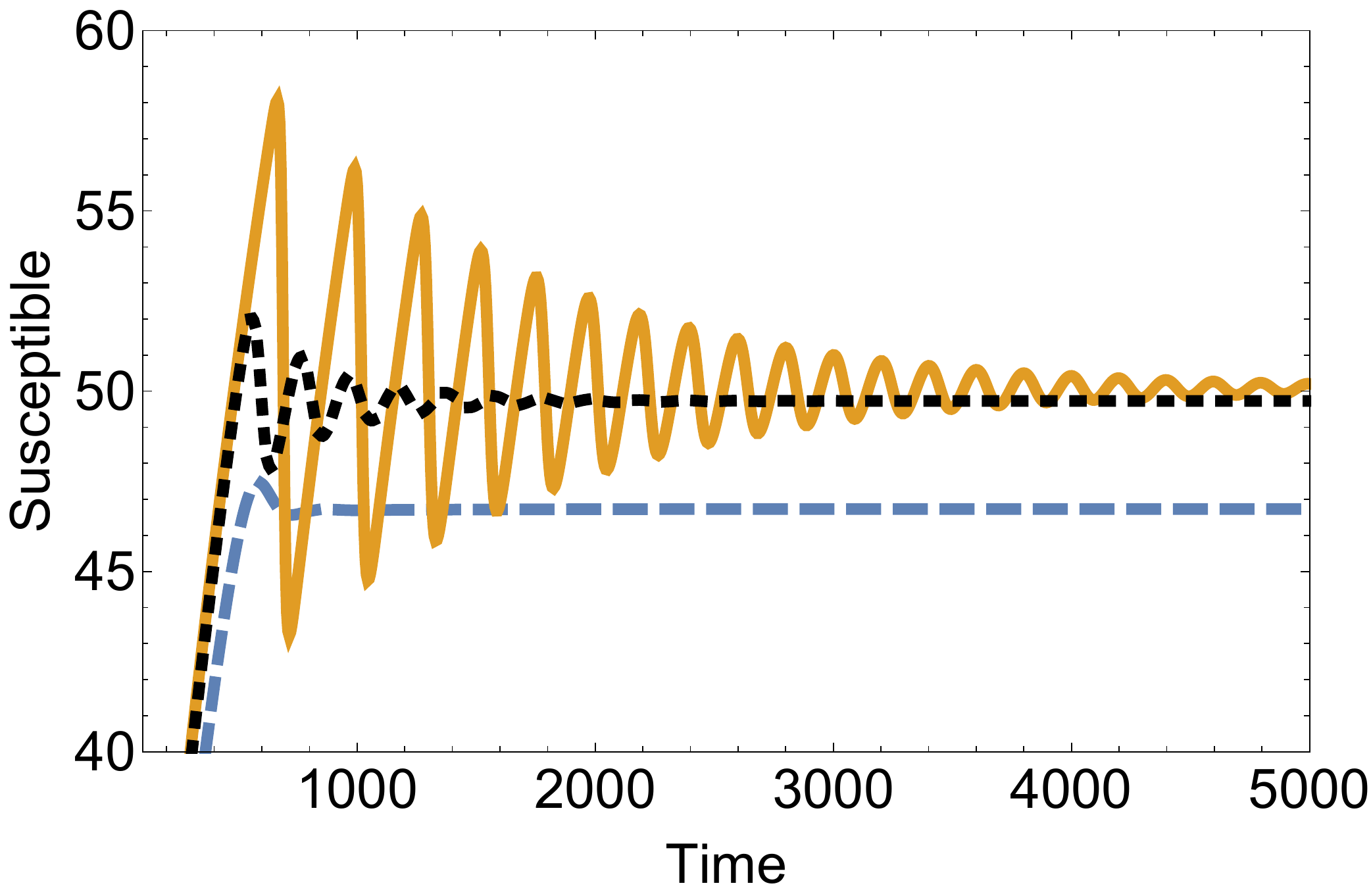}
\caption{Plot of susceptible versus time, after the initial relaxation time, for $\alpha=1$ (solid line), $\alpha=0.999$ (dotted line) and $\alpha=0.99$ (dashed line). The other parameters are $\lambda=0.1$, $\omega=0.02$, $\gamma=0.001$, and $\Delta t=0.05$.}
\label{fig_transient}
\end{center}
\end{figure}

For values of $\alpha$ very close to one the numerical solutions show an oscillatory approach towards the steady state. This can be seen in
Fig. (\ref{fig_transient}) for $\alpha=0.99$, $0.999$, and $1$. The oscillations are suppressed as $\alpha$ is decreased.

\section{Summary}
\label{sec_conclusion}
We have derived a fractional order recovery SIR model that differs from the classic SIR model by considering the case where the recovery from a disease does not follow an exponential distribution, but follows a distribution with a power law tail. The fractional order model is biologically motivated by the observation that in some disease processes, the longer a person is infectious the more likely they will remain infectious. The fractional order model permits both a disease free equilibrium state and an endemic equilibrium state. We have related the fractional order model to the generalised SIR models introduced by Kermack and McKendrick. The fractional order model that we have derived is different to {\em ad-hoc} fractional order epidemic models with Caputo fractional derivatives in time replacing standard derivatives in time.  Our work justifies an appropriate inclusion of fractional derivatives in SIR model equations.

We have also derived a discrete time fractional order recovery model that limits to the continuous time fractional order model as the time step goes to zero. The discrete time model was itself derived from a stochastic process and we have used this model to provide a stable numerical solver for the continuous time model. 
Our numerical solutions based on this discrete model show that the prevalence of infection and the peak levels of infection are both elevated by the fractional order derivative as it is varied further away from the integer order derivative in the classic SIR model.

\section*{Appendix A}
\label{sec_KMdeivation}
{\bf Derivation of the Kermack-McKendrick Model for Infecteds in a Age-Structured System}\newline

Let $i(a,t)$ denote the number of individuals who are infected at time $t$ and who have been infected for time $a$.
The total number of infected individuals at time $t$ is given by
\begin{equation}
\int_0^t i(a,t)\, da=\int_0^t \Phi(t,t')i(0,t')\, dt'\label{A1}
\end{equation}
where 
\begin{equation}
\Phi(t,t')=\phi(t-t')\theta(t,t')\label{A2}
\end{equation}
is the product of the probability of surviving death, $\theta(t,t')$, and of `surviving' the transition out of the next stage, $\phi(t-t')$.
If we differentiate Eq.(\ref{A1}) with respect to $t$ using Leibniz rule we have
\begin{equation}
\int_0^t\frac{\partial i}{\partial t}\, da+i(t,t)=\int_0^t\frac{\partial \Phi(t,t')}{\partial t}i(0,t')\, dt'
+\Phi(t,t)i(0,t)\label{A3}
\end{equation}
We also have
\begin{equation}
\int_0^t\frac{\partial i}{\partial a}\, da=i(t,t)-i(0,t).\label{A4}
\end{equation}
But
\begin{equation}
\Phi(t,t)=1
\end{equation}
so that if we add Eq.(\ref{A3}) and Eq.(\ref{A4}) then we obtain
\begin{equation}
\int_0^t\left(\frac{\partial i}{\partial t}+\, \frac{\partial i}{\partial a}\right)\, da=\int_0^t\frac{\partial \Phi(t,t')}{\partial t}i(0,t')\, dt'\label{A5}
\end{equation}
We can expand the partial derivative of $\Phi(t,t')$ using the product rule in Eq.(\ref{A2}) with
\begin{equation}
\frac{\partial\theta(t,t')}{\partial t}=-\gamma(t)\theta(t,t')
\end{equation}
and
\begin{equation}
\frac{\partial\phi(t-t')}{\partial t}=-\psi(t-t'),
\end{equation}
to obtain
\begin{equation}
\int_0^t\left(\frac{\partial i}{\partial t}+\, \frac{\partial i}{\partial a}\right)\, da=-\int_0^t\psi(t-t')\theta(t,t')i(0,t')\, dt'
-\int_0^t\gamma(t)\theta(t,t')\phi(t-t')i(0,t')\, dt'.\label{A5}
\end{equation}
Using Eq.(\ref{A2}) and then Eq.(\ref{A1}) in the second term on the RHS of Eq.(\ref{A5}) we now have
\begin{equation}
\int_0^t\left(\frac{\partial i}{\partial t}+\, \frac{\partial i}{\partial a}\right)\, da=-\int_0^t\psi(t-t')\theta(t,t')i(0,t')\, dt'
-\gamma(t)\int_0^t i(a,t)\, da.\label{A6}
\end{equation}
We now define
\begin{equation}
\beta(t-t')=\frac{\psi(t-t')}{\phi(t-t')}
\end{equation}
and use Eq.(\ref{A2}) in the first term on the RHS of Eq.(\ref{A6}) to obtain
\begin{equation}
\int_0^t\left(\frac{\partial i}{\partial t}+\, \frac{\partial i}{\partial a}\right)\, da=-\int_0^t\beta(t-t')\Phi(t,t')i(0,t')\, dt'
-\gamma(t)\int_0^t i(a,t)\, da.\label{A7}
\end{equation}
In the first integral on the RHS of Eq.(\ref{A7}) it is convenient to make a change of variables $t'=t-a$ with $dt'=-da$, then we have
\begin{equation}
\int_0^t\left(\frac{\partial i}{\partial t}+\, \frac{\partial i}{\partial a}\right)\, da=-\int_0^t\beta(a)\Phi(t,t-a)i(0,t-a)\, da
-\gamma(t)\int_0^t i(a,t)\, da.\label{A8}
\end{equation}
Finally we note that
\begin{equation}
i(a,t)=\Phi(t,t-a)i(0,t-a)
\end{equation}
so that Eq.(\ref{A8}) can be written as
\begin{equation}
\int_0^t\left(\frac{\partial i}{\partial t}+\, \frac{\partial i}{\partial a}\right)\, da=-\int_0^t\beta(a)i(a,t)\, da
-\gamma(t)\int_0^t i(a,t)\, da.\label{A9}
\end{equation}
This is consistent with Eq.(\ref{eq_KMi}) for the change in infectives in the age-structured Kermack McKendrick model.

\section*{Appendix B}
\label{sec_conttime}
{\bf Limits To Continuous Time}\newline

The discrete time fractional recovery SIR model can be shown to limit to
the frSIR model by identifying $t=n\Delta t$ and taking the limit $\Delta t \rightarrow 0$ with $r/\Delta t^{\alpha}$ finite. 
The continuous time equations can be obtained from the discrete time equations using Z star transform methods.
The $Z$ star transform of $Y(n)$ is given by
\begin{equation}
Z^*[Y(n)|s,\Delta t]=\sum_{n=0}^\infty Y(n) e^{-ns\Delta t}
\end{equation}
It follows that
\begin{eqnarray}
\Delta t Z^*[Y(n)|s,\Delta t]&=&\sum_{n=0}^\infty Y(n) e^{-ns\Delta t}\Delta t\\
&=&\sum_{n=0}^\infty \tilde Y(n\Delta t) e^{-ns\Delta t}\Delta t
\end{eqnarray}
where we have introduced $\tilde Y(t)$ as a function defined over a continuous variable $t$. 
We can now take the inverse Laplace transform from $s$ to $t$ 
\begin{equation}
\mathcal{L}^{-1}_s\left[\Delta t Z^*[Y(n)|s,\Delta t]\Bigg\vert t\right]
=\sum_{n=0}^\infty \tilde Y(n\Delta t) \delta(t-n\Delta t)\Delta t
\end{equation}
where $\delta(t)$ is the Dirac delta function.
Here, and in the following, we use the notation
$
\mathcal{L}^{-1}_s\left[ Y(s)\Bigg\vert t\right]
$ to denote the inverse Laplace transform from $s$ to $t$
and we use the notation
$
\mathcal{L}_t\left[Y(t)\Bigg\vert s\right]
$ to denote the Laplace transform from $t$ to $s$.

It is useful to define the function
\begin{equation}
\tilde Y(t|\Delta t)=\sum_{n=0}^\infty \tilde Y(n\Delta t) \delta(t-n\Delta t)\Delta t.\label{Yt}
\end{equation}
In a similar fashion we have
\begin{eqnarray}
\mathcal{L}^{-1}_s\left[\Delta t Z^*[Y(n-1)|s,\Delta t] \Bigg\vert t \right]
&=&\sum_{n=0}^\infty \tilde Y(n\Delta t) \delta(t-(n+1)\Delta t)\Delta t\\
&=&\sum_{n=0}^\infty \tilde Y((n-1)\Delta t)\delta(t-n\Delta t)\Delta t\\
&=&\tilde Y(t-\Delta t|\Delta t).
\end{eqnarray}
Note that, with $t'=n\Delta t$, in Eq.(\ref{Yt}), we have
\begin{eqnarray}
\lim_{\Delta t\to 0}\tilde Y(t|\Delta t)&=&\lim_{\Delta t\to 0}\sum_{n=0}^\infty \tilde Y(n\Delta t) \delta(t-n\Delta t)\Delta t\\
&=&\int_0^\infty \tilde Y(t')\delta(t-t')\, dt'\\
&=&\tilde Y(t).
\end{eqnarray}
This formally identifies
\begin{equation}
\tilde Y(t)=\lim_{\Delta t\to 0}
\mathcal{L}^{-1}_s\left[\Delta t Z^*[Y(n)|s,\Delta t] \Bigg\vert t\right],
\end{equation}
provided that the limit exists.

We further note the product rule
\begin{eqnarray}
& &\lim_{\Delta t\to 0}\sum_{n=0}^\infty \tilde X(n\Delta t)\tilde Y(n\Delta t)\delta(t-n\Delta t)\Delta t\nonumber\\
& &=\left(\lim_{\Delta t\to 0}\sum_{n=0}^\infty \tilde X(n\Delta t)\delta(t-n\Delta t)\Delta t\right)
\left(\lim_{\Delta t\to 0}\sum_{n=0}^\infty \tilde Y(n\Delta t)\delta(t-n\Delta t)\Delta t\right),\label{prodrule}
\end{eqnarray}
which equates to $\tilde X(t)\tilde Y(t)$ in each case, with $t'=n\Delta t$, provided that
both $\tilde X(t)$ and $\tilde Y(t)$ exist.

We now take the inverse Laplace transform of the  Z star transform of Eq.(\ref{eq_DI}) and multiply by $\frac{\Delta t}{\Delta t}$ to write
\begin{equation}
\begin{split}
\sum_{n=0}^\infty & \frac{\tilde I(n\Delta t)-\tilde I((n-1)\Delta t)}{\Delta t}\delta(t-n\Delta t)\Delta t\\
=&\sum_{n=0}^\infty \frac{\tilde\omega(n\Delta t)}{\Delta t}  \tilde S((n-1)\Delta t)\tilde I((n-1)\Delta t)\delta(t-n\Delta t)\Delta t\\
& -\frac{r}{\Delta t}\sum_{n=0}^\infty\tilde \theta(n\Delta t,0)\left(\sum_{k=0}^{n-1}\tilde\kappa((n-k)\Delta t)\frac{\tilde I(k\Delta t)-\tilde I_0(k\delta t)}{\tilde\theta(k\Delta t,0)}\right)\delta(t-n\Delta t)\Delta t\\
& -\sum_{n=0}^\infty\frac{\tilde\gamma(n\Delta t)}{\Delta t} \left(\tilde I((n-1)\Delta t)-\tilde I_0((n-1)\delta t)\right)\delta(t-n\Delta t)\Delta t\\
&+\sum_{n=0}^\infty \frac{\tilde I_0(n\Delta t)-\tilde I_0((n-1)\Delta t)}{\Delta t}\delta(t-n\Delta t)\Delta t \label{horror1}
\end{split}
\end{equation}

We now take the continuous time limit of Eq.(\ref{horror1}) using $t'=n\Delta t$ and the product rule in Eq.(\ref{prodrule}),
to obtain
\begin{equation}
\begin{split}
\int_0^\infty &\left( \lim_{\Delta t\to 0}\frac{\tilde I(t')-\tilde I(t'-\Delta t)}{\Delta t}\right)\delta(t-t')\, dt'=\\
& \int_0^\infty \hat\omega(t')  \tilde S(t')\tilde I(t')\delta(t-t')\, dt'\\
& -\left(\int_0^\infty\tilde\theta(t',0)\delta(t-t')\, dt'\right)\left(\lim_{\Delta t\to 0}\frac{r}{\Delta t}\sum_{n=0}^\infty\left(\sum_{k=0}^{n-1}\tilde\kappa((n-k)\Delta t)\frac{\tilde I(k\Delta t)-\tilde I_0(k\Delta t)}{\tilde\theta(k\Delta t,0)}\right)\delta(t-n\Delta t)\Delta t\right)\\
& -\int_0^\infty\hat\gamma(t')\left(\tilde I(t')-\tilde I_0(t')\right)\delta(t-t')\, dt'\\
&+\int_0^\infty \left( \lim_{\Delta t\to 0}\frac{\tilde I_0(t')-\tilde I_0(t'-\Delta t)}{\Delta t}\right)\delta(t-t')\, dt'\label{lesshorror}
\end{split}
\end{equation}
where we have defined continuous time rate parameters
\begin{equation}
\hat \omega(t')=\lim_{\Delta t\to 0}\frac{\tilde\omega(n\Delta t)}{\Delta t}\label{omegat}
\end{equation}
and
\begin{equation}
\hat \gamma(t')=\lim_{\Delta t\to 0}\frac{\tilde\gamma(n\Delta t)}{\Delta t}.
\end{equation}
Equation (\ref{lesshorror}) simplifies further to
\begin{equation}
\begin{split}
\frac{d\tilde I(t)}{dt}&= \hat \omega(t)\tilde S(t)\tilde I(t)-\hat\gamma(t)\left(\tilde I(t)-\tilde I_0(t)\right)+\frac{d\tilde I_0(t)}{dt}\\
& -\tilde\theta(t,0)\left(\lim_{\Delta t\to 0} \frac{r}{\Delta t}\sum_{n=0}^\infty\left(\sum_{k=0}^{n-1}\tilde\kappa((n-k)\Delta t)\frac{\tilde I(k\Delta t)-\tilde I_0(k\Delta t)}{\tilde\theta(k\Delta t,0)}\right)\delta(t-n\Delta t)\Delta t\right)\label{better}
\end{split}
\end{equation}

The further reduction of this equation depends on the specific form of the memory kernel $\kappa(n)$.
In the case of a jump at each time step the memory kernel is,
\begin{equation}
\kappa(n)=\delta_{n,1}.
\end{equation}
In this case we can perform the sum over $k$ explicitly in Eq. (\ref{better}) to arrive at
\begin{eqnarray}
\begin{split}
\frac{d\tilde I(t)}{dt}&= \hat \omega(t)\tilde S(t)\tilde I(t)-\hat\gamma(t)\left(\tilde I(t)-\tilde I_0(t)\right)+\frac{d\tilde I_0(t)}{dt}\\
&-\tilde\theta(t,0)\left(\lim_{\Delta t\to 0} \frac{r}{\Delta t}\sum_{n=0}^\infty\frac{\tilde I((n-1)\Delta t)-\tilde I_0((n-1)\Delta t)}{\tilde\theta((n-1)\Delta t,0)}\delta(t-n\Delta t)\Delta t\right)
\end{split}
\end{eqnarray}
In order for the continuous time limit of the above equation to exist we define
\begin{equation}
\mu=\lim_{\Delta t\to 0}\frac{r}{\Delta t}.
\end{equation}
Note that $r$ is a free parameter in the range $[0,1]$ and hence $\mu$ is only well defined in this limit if we take $r$ to be a function of $\Delta t$. 
With this definition of $\mu$ we can now perform the limit $\delta t\to 0$ to obtain the continuous time equation
\begin{equation}
\frac{d\tilde I(t)}{dt}
=\hat\omega(t)
\tilde S(t)\tilde I(t)
-\mu(\tilde I(t)-\tilde I_0(t))-\hat\gamma(t)(\tilde I(t)-\tilde I_0(t))+\frac{d\tilde I_0(t)}{dt}.
\end{equation} 
This further simplifies to
\begin{equation}
\frac{d\tilde I}{dt}
=\hat\omega(t)
\tilde S(t)\tilde I(t)
-\mu\tilde I(t)-\hat\gamma(t)\tilde I(t)\label{Icont3}
\end{equation} 
Equation (\ref{Icont3}) recovers the corresponding equation in the classic SIR model.

We now consider the continuous time limit of Eq.(\ref{better}) with the Sibuya memory kernel, given by Eq.(\ref{kappa}). First we simplify the double sum in Eq.(\ref{better})  using Laplace transforms and Z star transforms
as follows:
\begin{eqnarray}
& &\lim_{\Delta t\to 0}\frac{r}{\Delta t}\sum_{n=0}^\infty\left(\sum_{k=0}^{n-1}\tilde\kappa((n-k)\Delta t)\frac{\tilde I(k\Delta t)-\tilde I_0(k\Delta t)}{\tilde\theta(k\Delta t,0)}\right)\delta(t-n\Delta t)\Delta t\nonumber\\
& &=
\mathcal{L}_s^{-1}\left[\mathcal{L}_t\left[\lim_{\Delta t\to 0}\frac{r}{\Delta t}
\sum_{n=0}^\infty\left(\sum_{k=0}^{n-1}\tilde\kappa((n-k)\Delta t)\frac{\tilde I(k\Delta t)-\tilde I_0(k\Delta t)}{\tilde\theta(k\Delta t,0)}\right)\delta(t-n\Delta t)\Delta t\Bigg\vert s\right]\Bigg\vert t\right]\nonumber\\
& &=\mathcal{L}_s^{-1}\left[\lim_{\Delta t\to 0}\frac{r}{\Delta t}
\sum_{n=0}^\infty\left(\sum_{k=0}^{n-1}\tilde\kappa((n-k)\Delta t)\frac{\tilde I(k\Delta t)-\tilde I_0(k\Delta t)}{\tilde\theta(k\Delta t,0)}\right)e^{-sn\Delta t}\Delta t\Bigg\vert t\right]\nonumber\\
& &=\mathcal{L}_s^{-1}\left[\lim_{\Delta t\to 0}\frac{r}{\Delta t} Z^*[\sum_{k=0}^{n-1}\kappa(n-k)\frac{I(k)-I_0(k)}{\theta(k,0)}| s,\Delta t ]\Delta t\Bigg\vert t\right]\nonumber\\
& &=
\mathcal{L}_s^{-1}\left[ \lim_{\Delta t\to 0}\frac{r}{\Delta t}Z^*[\kappa(n)| s,\Delta t]
Z^{*}[\frac{I(n)-I_0(k)}{\theta(n,0)}| s,\Delta t]\Delta t\Bigg\vert t\right].\label{monster}
\end{eqnarray}
The last line in the above follows from the convolution theorem for Z star transforms.

To proceed further we use the Z transform of the Sibuya memory kernel in Eq.(\ref{Zkappa}) to write
\begin{eqnarray}
Z^*[\kappa(n)| s,\Delta t]
&=&\left[ ((1-e^{-s\Delta t})^{1-\alpha}-(1-e^{-s\Delta t}))\right]\\
&\approx& (s\Delta t)^{1-\alpha}+o(s\Delta t).
\end{eqnarray}
The result in Eq.(\ref{monster}) can now be written as
\begin{eqnarray}
& &\lim_{\Delta t\to 0}\frac{r}{\Delta t}\sum_{n=0}^\infty\left(\sum_{k=0}^{n-1}\tilde\kappa((n-k)\Delta t)\frac{\tilde I(k\Delta t)-\tilde I_0(k\Delta t)}{\tilde\theta(k\Delta t,0)}\right)\delta(t-n\Delta t)\Delta t\nonumber\\
& &=
\mathcal{L}_s^{-1}\left[ \lim_{\Delta t\to 0}\frac{r}{\Delta t^\alpha} s^{1-\alpha}
\sum_{n=0}^\infty \frac{\tilde I(n\Delta t)-\tilde I_0(n\Delta t)}{\tilde \theta(n\Delta t,0)}e^{-sn\Delta t}\Delta t\Bigg\vert t\right]\nonumber\\
&=&
\mu \mathcal{L}_s^{-1}\left[ s^{1-\alpha}\int_0^\infty \frac{\tilde I(t)-\tilde I_0(t)}{\tilde \theta(t,0)}e^{-st}\, dt\Bigg\vert t\right]\nonumber\\
&=&
\mu \mathcal{L}_s^{-1}\left[ s^{1-\alpha} \mathcal{L}_t \left[\frac{\tilde I(t)-\tilde I_0(t)}{\tilde \theta(t,0)}\Bigg\vert s\right]\Bigg\vert t\right],\label{result}
\end{eqnarray}
where
\begin{equation}
\mu=\lim_{\Delta t\to 0}\frac{r}{\Delta t^\alpha}.
\end{equation}

Finally we substitute the result of Eq.(\ref{result}) into Eq.(\ref{better}), and use
the known result \cite{P1999}
\begin{equation}
\mathcal{L}_t\left[\, _0D_t^{1-\alpha} Y(t)\Bigg\vert s\right]
= s^{1-\alpha} \mathcal{L}_t\left[Y(t)\Bigg\vert s\right]
\end{equation}
 to invert the Laplace transform and obtain
\begin{equation}
\frac{d\tilde I(t)}{dt}
=\hat{\tilde{\omega}}(t)
\tilde S(t)\tilde I(t)
-\mu\hat{\tilde\theta}(t,0)
\, _0D_t^{1-\alpha}\left(\frac{\tilde I(t)-\tilde I_0(t)}{\hat{\tilde\theta}(t,0)}\right)
-\hat{\tilde\gamma}(t)\left(\tilde I(t)-\tilde I_0(t)\right)+\frac{d\tilde I_0(t)}{dt}\label{Icont4}
\end{equation} 
Equation (\ref{Icont4}) recovers the continuous time frSIR model equation.

Note that in order for the continuous time limit of the frSIR model equation to exist we defined
\begin{equation}
\mu=\lim_{\Delta t\to 0}\frac{r}{\Delta t^\alpha}\label{rmu}
\end{equation}
which requires $r\in[0,1]$ to be a function of $\Delta t$.
This is important for numerical simulations based on this DTRW method where
we take
$r=\mu \Delta t^\alpha$ and then the requirement that $r\in[0,1]$ places restrictions
on $\Delta t$ for given $\alpha$ and $\mu$.

\begin{acknowledgements}
This work was supported by the Australian Research Council  (DP130100595, DP140101193).
\end{acknowledgements}


\end{document}